\documentclass[aps,floatfix]{revtex4}

\usepackage{amsmath,amsthm,amsfonts,amssymb,times,graphicx}
\newtheorem{theorem}{Theorem}
\newtheorem{proposition}[theorem]{Proposition}

\newtheorem{corollary}[theorem]{Corollary}
\newtheorem{definition}[theorem]{Definition}
\newtheorem{remark}[theorem]{Remark}

\newcommand{\ket}[1]{\mbox{$| #1 \rangle$}}
\newcommand{\braket}[2]{\mbox{$\langle #1  | #2 \rangle$}}

\def\ZZ{\mathbb{Z}}

\def\CC{\mathbb{C}}

\def\Map{\textbf{Map}}

\begin{document}

\title{Efficient 2-designs from bases exist}

\author{Gary McConnell and David Gross}

\affiliation{
	Institute for Mathematical Sciences, Imperial College London, London
	SW7 2BW, UK
}
\affiliation{
	QOLS, Blackett Laboratory, Imperial College London, London SW7 2BW,
	UK
}

%\email{davidg@ic.ac.uk}

\date{\today}

\begin{abstract}
We show that in a complex $d$-dimensional vector space, one can find
$O(d)$ bases whose elements form a 2-design. Such vector sets
generalize the notion of a maximal collection of mutually unbiased
bases (MUBs). MUBs have manifold applications in quantum information
theory (e.g.\ in state tomography, cloning, or cryptography) --
however it is suspected that maximal sets exist only in prime-power
dimensions. Our construction offers an efficient alternative for
general dimensions. The findings are based on a framework recently
established in [A.\ Roy and A.\ Scott, J.\ Math.\ Phys.\ {\bf 48},
072110 (2007)], which reduces the construction of such bases to the
combinatorial problem of finding certain highly nonlinear functions
between abelian groups.
\end{abstract}

%\pacs{}

\maketitle

\section{Introduction}

Two bases $\{\ket{e_i}\}_{i=1,\dots,d}$ and
$\{\ket{f_i}\}_{i=1,\dots,d}$ in a $d$-dimensional Hilbert space are
called \emph{mutually unbiased} if $|\braket{e_i}{f_j}|^2=1/d$ for
every $i,j$. It has been shown that there can exist no more than $d+1$ such
bases in $\CC^d$, and, conversely, that this number can be attained
whenever $d$ is the power of a prime \cite{mubs}. It is intuitive that
MUBs are advantageous for quantum state tomography, as measurements in
unbiased bases reveal ``maximally complementary'' information about the
measured state.

One can make rigorous the intuition that MUBs are ``evenly spread out'' in
state space, by observing that the elements of a maximal
collection of MUBs form a \emph{complex projective 2-design}
\cite{designs,koenig,klappenecker,unitary,emerson}.  Roughly
speaking, a set of vectors $\mathcal{D}$ is called a $t$-design, if
the average of every $t$th order polynomial $f$ over the unit sphere
in $\CC^d$ equals the average of $f$ over $\mathcal{D}$ (see
Definition \ref{def:weighted} below). Several of the advantageous
properties of MUBs follow directly from this feature: e.g.\ a simple
formula for state reconstruction in terms of measurement outcomes or
their optimality in certain cloning protocols \cite{aidan}.

A considerable amount of research has gone into the problem of
determining $MUB(d)$, the number of MUBs in dimension $d$ \cite{mubs}.
Little is known about $MUB(d)$ when $d$ is not a power of a
prime -- however, there is some evidence for the fact that
$MUB(d)<d+1$ in these cases \cite{grassl,nogeneralizations,aschbacher}.
While determining $MUB(d)$ is certainly an important mathematical
problem, it may not be the most pertinent question to ask from the point of
view of quantum state tomography, as only maximal sets of MUBs can be
used for this purpose. So it is timely to look for a ``second best''
alternative to maximal sets of MUBs.

Therefore, in \cite{aidan} it was proposed that the problem be
approached from a different direction. The authors examine the
quantity $M(d)$, defined as the number of bases one needs in dimension
$d$ in order to form a 2-design. The number $M(d)$ equals $d+1$ if and
only if there is a complete set of MUBs in $d$ dimensions. In general,
$M(d)>d+1$, but whenever $M(d)$ is reasonably small, such sets of
bases can serve as a good substitute for MUBs \cite{aidan}.  We call a
2-design of this kind \emph{efficient} if it consists of $O(d)$ bases.

It was shown in \cite{aidan} that $M(d)\leq \frac34
(d-1)^2$.  Here, we improve their results by constructing weighted
complex projective 2-designs from roughly $2(d+\sqrt d)$ bases for odd $d$ and
$3(d+\sqrt d)$ bases in even dimensions.

\section{Definitions and previous results}

Let $f$ be a homogeneous polynomial of order $t$ in $2d$ variables. We
can regard $f$ as a function on $\CC^d$ by evaluating it on
coordinates (with respect to an arbitrary fixed basis) and their
complex conjugates:
$f(\ket\psi)=f(\psi_1,\dots,\psi_d;\bar{\psi_1},\dots,\bar{\psi_d})$.
The set of such polynomials is denoted by $\operatorname{Hom}(t,t)$.

\begin{definition}[Weighted 2-designs]\label{def:weighted}
	Let $\mathcal{D}$ be a set of normalized vectors in $\CC^d$ and $w:
	\mathcal{D}\to[0,1]$ a normalized weight function.  The set
	$\mathcal{D}$ together with the weights $w$ is a \emph{weighted
	complex projective 2-design} if for all
	$f\in\operatorname{Hom}(2,2)$ the relation
	\begin{equation}\label{eqn:def}
		\sum_{x\in\mathcal{D}} w(x) f(x) = \int_{\CC P^{d-1}} f(x) dx
	\end{equation}
	holds.
\end{definition}

The integral on the right hand side of (\ref{eqn:def}) is understood to
be taken with respect to the Haar measure on $\CC P^{d-1}$. We will make use
of a combinatorial construction for weighted 2-designs introduced in
\cite{aidan}. To this end:

\begin{definition}[Differential 1-uniformity \cite{aidan}]\label{d1udefn}
	Let $A,B$ be finite abelian groups. The function $f: A\to B$ is
	\emph{differentially 1-uniform} (d1u) if the equation
	\begin{equation}\label{eqn:eqn}
		f(x+a)-f(x)=b
	\end{equation}
	has at most one solution in $x$ for every $(a,b)\neq (0,0)$.
\end{definition}

Differentially 1-uniform functions are related to \emph{highly
non-linear functions}, which have been the subject of research in
combinatorics and cryptography \cite{highlynl}. 
%Note however that in
%this field one generally assumes $|A|\geq |B|$, while we will be
%concerned with the case $|B|\geq |A|$. Hence the results are not
%directly comparable.

\begin{theorem}[2-designs from d1u functions \cite{aidan}]
	If $f:A\to B$ is d1u, then there is a weighted complex projective
	2-design formed from $|B|+1$ bases in dimension $d=|A|$.
\end{theorem}

Hence the challenge is to construct d1u functions from general $A$ to
some $B$ which is as small as possible. Our particular construction below
makes use of d1u functions with \emph{cyclic domain} $A=\ZZ/d\ZZ$.

For any positive integer $d$,
denote by $\mathbf{C}(d)$ the smallest cardinality of an abelian group $B$ such that
there exists a d1u function $f: \ZZ/d\ZZ \to B$.  The following theorem
summarizes the relevant results of \cite{aidan}.

\begin{theorem}[Known bounds on $\mathbf{C}(d)$ \cite{aidan}]
	\label{knownd1u}
	With notation as above:
	\begin{enumerate}
		\item
		If $d$ is an odd prime power, then $\mathbf{C}(d)=d$ (which is optimal).

		\item
		For $d=p^k-1$, where $p$ is an arbitrary prime number and $k$ is any positive integer,
		we have $\mathbf{C}(d)\leq d+1$.

		\item
		For general $d$, $\mathbf{C}(d)\leq \frac34(d-1)^2$.
	\end{enumerate}
\end{theorem}

\section{An $O(d)$ bound for $\mathbf{C}(d)$}

We aim to improve the bounds of Theorem \ref{knownd1u}. The essence of
the result is that $\mathbf{C}(d)$ is linear in $d$:

\begin{theorem}
\label{Odthm}
$\mathbf{C}(d) = O(d).$
\end{theorem}

More precisely, let $d$ be any integer $\geq2$.
Let $q_d$ denote the smallest integer $\geq d-1$ such that there exists a d1u
function $\ZZ/q_d\ZZ \to B$ whose codomain $B$ is of minimal order $|B| = \mathbf{C}(q_d)$,
among all such integers and d1u functions.

%Suppose that $q$ is
%an integer $\geq d-1$ such that there exists an abelian group $B_q$
%and a d1u function $\ZZ/q\ZZ\to B_q$. Among all such $q$ there is a
%$B_q$ whose order is minimal (equal to $\mathbf{C}(q)$).  Define $q_d$ to
%be this value of $q$.
For example, if $d-1$ is an odd prime,
then by Theorem \ref{knownd1u} we can take $q_d=d-1,\ B=\ZZ/q\ZZ$ and
clearly then $|B|=q$ will be minimal.

The key result of this paper is as
follows.

\begin{theorem}
\label{mainthm}
Let $d$ be any integer $\geq2$ and define $q_d$ as above.
\begin{enumerate}
	\item
	If $d$ is odd then
	$\mathbf{C}(d)\leq 2\,\mathbf{C}(q_d);$

	\item
	and if $d$ is even,
	$\mathbf{C}(d)\leq 3\,\mathbf{C}(q_d).$
\end{enumerate}
\end{theorem}

By taking $q_d$ to be the smallest prime greater than or equal to $d$, we
get the following explicit asymptotic bounds:

\begin{corollary}\label{core}
Let $d$ be as above and let $\theta=0.525$ \cite{bakerHarman}. For
$d$ large enough, we have that
\begin{enumerate}
	\item
	for $d$ odd,
    $\mathbf{C}(d)\leq 2(d+d^\theta);$

	\item
	and similarly for $d$ even,
	$\mathbf{C}(d)\leq 3(d+d^\theta).$
\end{enumerate}
\end{corollary}

We prove these results by constructing explicit functions
from $\ZZ/d\ZZ$ into groups of the sizes shown.

\subsection{Differentials and group homomorphisms}

Let $A, B$ be two arbitrary finite abelian groups, written additively.
We shall assume that $|B|\geq|A|\geq 2$.  Let $\Map(A,B)$
denote the set of all functions between $A$ and $B$, which itself is a
finite abelian group under pointwise addition.

Given any $a\in A$ and $f:A\to B$, define the differential
operator $D_a: \Map(A,B) \to \Map(A,B)$ by $$D_a(f)(x) = f(a+x) -
f(x),$$ for any $x$ in $A$.  In this terminology, Definition
\ref{d1udefn} may be rephrased thus: the function $f$ is d1u if for all
non-zero $a\in A$, the vector $\big(D_a(f)(x)\big)$ contains no
repeated values (here we fix an ordering of the elements $x\in A$).
This makes precise the somewhat loose notion that $f$ is d1u if it is
as ``far from being a homomorphism as possible''.  Indeed, yet another
equivalent formulation of the condition that a function $f$ be d1u is
that its second differentials $D_{a_1}D_{a_2}f$ be nowhere-vanishing
for all $a_1,a_2\in A\setminus\{0\}$.

There is a symmetry relation among the vectors $D_a(f)$ of
differentials which follows from the identity:
\begin{equation}
\label{wraparound}
D_a(f)(x) = -D_{-a}(f)(a+x),
\end{equation}
for all $a,x\in A$.  As a practical matter therefore, to check if a
function $f$ is d1u, it suffices to check ``the first half'' of the vectors
$D_a(f)$.  In addition, the following useful identity holds:
\begin{equation}
\label{DiIterate}
D_{ra}(f)(x) = \sum_{i=0}^{r-1} D_a(f)(ia+x),
\end{equation}
for all $r\in\ZZ$, $x\in A$ and for all nonzero $a\in A$.
If $A$ is cyclic then each vector $D_af$ is easily determined using
(\ref{DiIterate}) from the one generating vector $D_1f$.  This
obviously also holds within the cyclic subgroups of a general abelian
group $A$.

\subsection{The construction}

We now present a new class of d1u functions which improves the bound
in (3) of Theorem \ref{knownd1u}.  Henceforth we assume $A=\ZZ/d\ZZ$.
The aim is to find $n$ as small as possible such that there exists a
group $B$ of order $n$ and a d1u function $f:A\to B$.

So let $d$ be any integer $\geq 2$.  Let $p$ be the least prime which is coprime to $d$.  Let $q$ be any integer $\geq d-1$ such that there exists a finite abelian group $G$ and a d1u function $\phi:\ZZ/q\ZZ\to G$.  For example, by Theorem \ref{knownd1u} we may take $q$ to be any odd prime $\pi\geq d-1$, or else $q=\pi^k+1$ for $\pi$ any prime, $k\geq1$ (see \cite{aidan} \S4 for the actual
functions $\phi$ in this case).

We write $\phi_d$ for the `restriction' of $\phi$ to $\ZZ/d\ZZ$, viz.:
$$\phi_d(x) = \phi(x),\ 0\leq x\leq d-1,$$
where it is understood that $\phi_d$ is defined modulo $d$ and where we set $\phi_d(d-1)=\phi(q)=\phi(0)$ in the case $q=d-1$.

Then we are able to construct examples of d1u functions as follows.
We are grateful to Aidan Roy for pointing out the neat form in which our original (less general) construction has been rephrased below.

\begin{proposition}
\label{dlogd}
Let $d,p,q,G,\phi_d$ be as above.  Define $f: \ZZ/d\ZZ \to \ZZ/p\ZZ\times G$ by
\begin{equation}
\label{fdef}
f(i) = (i, \phi_d(i)),
\end{equation}
for $0\leq i\leq d-1$.  Then $f$ is d1u.
\end{proposition}

\begin{proof}
\textbf{Case (i): $q\geq d$.}
Fix $a\in\ZZ/d\ZZ$.  Observe that for every $x\in \ZZ/d\ZZ$ with $0\leq a+x\leq d-1$:
$$D_a\phi_d(x) = \phi_d(a+x)-\phi_d(x) = \phi(a+x)-\phi(x) = D_a\phi(x),$$
(where we always write $a$ and $x$ as the smallest positive integers
representing their respective congruence classes modulo $d$ or $q$ as
the case may be).  Since $\phi$ is d1u, we know therefore that the values $D_a\phi_d(x)$ are distinct in $G$ as $x$ runs from $0$ to $d-1-a$.  Hence the same must be true of the $D_af(x)$ for such $x$ because of the identity
$$D_af(x) = (D_a(x),\ D_a\phi_d(x)).$$
\begin{remark}
\label{wrap}
	Note that one must take some care with this function $D_a(x)$ when
	$a+x\geq d$: the shift function $x\mapsto a+x$ operates modulo
	$d$ (not $q$) and so in fact the congruence class of $a+x$ is $a+x-d$ for the
	purposes of evaluating these differentials.  For $0\leq x\leq d-1-a$
	there is no ambiguity and $D_a(x)$ is just equal to $a$.
\end{remark}
So it remains to show that the values of $D_af(x)$ for $d-a\leq x\leq d-1$ are distinct from one another, and that they do not coincide with any of the values just described for $0\leq x\leq d-1-a$.  This latter point follows from the remark above, since $p$ is coprime to $d$ and so for $a+x\geq d$:
$$D_a(x) = a+x-d-x = a-d \ncong a \mod p.$$
We have reduced the problem to the assertion that for $d-a\leq x\leq d-1$, the $D_a\phi_d(x)$ are distinct.  But it follows from the definitions of $\phi$ and $\phi_d$ that
$$D_a\phi_d(x) = \phi_d(a+x)-\phi_d(x) = \phi(a+x+q-d)-\phi(x) = D_{a+q-d}\phi(x),$$
which again by the choice of $\phi$ as a d1u function, cannot have repeated values inside $G$.

\textbf{Case (ii): $q = d-1$.}
The proof is almost identical to that for case (i), the only added complication being that one has to consider the value
$$D_a\phi_d(d-1-a) = \phi_d(d-1) - \phi_d(d-1-a) = \phi(0)-\phi(q-a)$$
which arises when $x=d-1-a(=q-a)$, and to show that it does not already exist in the set of $D_a\phi(x)$ for $0\leq x\leq q-a-1$.  But
$$\phi(0)-\phi(q-a) = D_a\phi(q-a),$$
which by the fact that $\phi$ is d1u cannot lie in the set described.
\end{proof}

So for odd $d$, we can choose $q$ to be prime and employ
Theorem \ref{knownd1u} to obtain an upper bound for $\mathbf{C}(d)$ of
twice the smallest prime $\geq d-1$. By Chebyshev's
theorem, there is always a prime between $d$ and $2d$, so
$\mathbf{C}(d)\leq 4 d$. Making use of more elaborate bounds on the worst
case gap between two consecutive primes \cite{bakerHarman}, we
obtain the result advertised in Corollary \ref{core}.

However, for even $d$ we are constrained to around $3d$ at best; and
for dimensions where $d$ is divisible by $3\times 2$, it is often much
worse.  For example if $d=30030$, then the best bound given by this
construction is $30029\times17=510510$. So we now provide a tighter
bound for even values of $d$.

\begin{proposition}
\label{evens}
Let $d,q,G,\phi,\phi_d$ be as above, with $d$ even.  Let a `flag' function $\theta:\ZZ/d\ZZ\to\ZZ/3\ZZ$ be defined by $\theta(x) = 0$ for $0\leq x\leq d/2-1$ and $\theta(x) = 1$ for $d/2\leq x\leq d-1$.  Then
\begin{equation}
%\label{fdef}
f: \ZZ/d\ZZ \to \ZZ/3\ZZ\times G\ :\ f(i) = (\theta(i), \phi_d(i))
\end{equation}
is d1u.
\end{proposition}

\begin{proof}
It is clear from the structure of the function $f$ that we may rely
almost completely upon the previous proof, bearing in mind Remark
\ref{wrap}.  By equation \ref{wraparound} we need only focus on
$a$ in the range $1\leq a\leq d/2$.  Then all we need to observe is
that in `crossing the $a+x=d$ threshold', the functions $D_af$ switch
their flag $D_a\theta(x)$ to $-1$ from $0$ (or from $+1$ for
$d/2-a\leq x\leq d/2-1$),  hence ensuring that the sets
$$D_af(x),\ 0\leq x\leq d-1-a$$
and
$$D_af(y) = (-1,D_{a+q-d}\phi(y)),\ d-a\leq y\leq d-1$$
remain disjoint.  Note that a similar observation to the one in Proposition \ref{dlogd} takes care of the case $q=d-1$.
\end{proof}

This ends the proof of Theorem \ref{mainthm}, and by extension of Theorem \ref{Odthm}.

\section{Computer findings in low dimensions}

The results presented above give solutions which are within a
multiplicative constant of the theoretical optimum $\mathbf{C}(d)=d$.
Still, computer searches reveal that better d1u functions
are very likely to exist in general -- at least whenever $d$ is
neither an odd prime power nor of the form
$p^k-1$ for prime $p$.  The first three numbers which are not of this form are $d=14,20$ and
$21$. The table below compares recent computer findings of Andrew Scott (private communication)
in these dimensions, with the systematic methods of the present note.
\begin{equation}
	\begin{tabular}{c|ccc}
		$d$ & 14 & 20 & 21 \\
		\hline
		Systematic: $\mathbf{C}(d)\leq$ & 39 & 57 & 46 \\
		Computer: $\mathbf{C}(d)\leq$ & 20 & 32 & 37
	\end{tabular}
\end{equation}
Scott has further shown by exhaustive search methods that no d1u
function exists from $\ZZ/14\ZZ$ into any abelian group of order less
than $20$.

\section{Acknowledgments}

We would like to thank Andrew Scott and Aidan Roy for their help in orienting this work
and Terry Rudolph for many thought-provoking conversations.

This work has been supported by QAP, the EURYI Award of J.\
Eisert, and the QIP-IRC.

\end{document}